\documentclass[11pt,preprint,superscriptaddress,aps,showkeys,nofootinbib]{revtex4}
\usepackage{amssymb,amsmath,amsfonts}
\usepackage{graphicx}
\usepackage{eepic,epsfig}
\usepackage{indentfirst}
\usepackage[utf8]{inputenc}
\usepackage[T1]{fontenc}

1.60cm \makeatletter \@addtoreset{equation}{section}

\makeatother

\begin{document}
\title{Fermionic Casimir effect in a field theory model with Lorentz symmetry violation}
\author{M. B. Cruz}
\affiliation{Departamento de F\'{\i}sica, Universidade Federal da Para\'{\i}ba\\
	Caixa Postal 5008, 58051-970, Jo\~ao Pessoa, Para\'{\i}ba, Brazil}
\email{messiasdebritocruz@gmail.com, emello, petrov@fisica.ufpb.br}
\author{E. R. Bezerra de Mello}
\affiliation{Departamento de F\'{\i}sica, Universidade Federal da Para\'{\i}ba\\
	Caixa Postal 5008, 58051-970, Jo\~ao Pessoa, Para\'{\i}ba, Brazil}
\email{messiasdebritocruz@gmail.com,  emello, petrov@fisica.ufpb.br}
\author{A. Yu. Petrov}
\affiliation{Departamento de F\'{\i}sica, Universidade Federal da Para\'{\i}ba\\
	Caixa Postal 5008, 58051-970, Jo\~ao Pessoa, Para\'{\i}ba, Brazil}
\email{messiasdebritocruz@gmail.com,  emello, petrov@fisica.ufpb.br}


\begin{abstract}
In this paper, we evaluate the Casimir energy and pressure for a massive fermionic field confined in the region between two parallel plates. In order to implement this confinement we impose  the standard  MIT bag boundary on the plates  for the fermionic field.  In this  paper we consider a quantum field theory model with a CPT even, aether-like Lorentz symmetry violation. It turns out that the fermionic Casimir energy and pressure depend on the direction of the constant vector that implements the Lorentz symmetry breaking.
\end{abstract}
\keywords{Lorentz symmetry breaking, Fermionic fields, Casimir effect, MIT bag model.}

\maketitle

\newpage
\section{Introduction}

The shift in the zero-point fluctuations of quantum fields  submitted to boundary conditions,  gives rise to macroscopically observable forces between material bodies, the so-called Casimir forces. The Casimir effect was proposed by H. B. Casimir in 1948 \cite{Casimir:1948dh}, and confirmed ten years later in the experiments  performed by Sparnaay \cite{Sparnaay:1958wg} who considered two parallel plates configuration; moreover, Blokland and Overbeek \cite{van1978phgm} have studied a plane surface opposing to a spherical one. More recently, these measurements have been performed using torsion balances \cite{lamoreaux1997sk}, atomic force microscopes \cite{Mohideen:1998}, and high precision capacitance bridges \cite{chan2001hb} (for a general review of the Casimir effect, see also \cite{Bordag,Milton}; some recent  discussions of different issues related to boundary conditions within the Casimir effect can be also found in \cite{Asorey,Asorey1,Munoz}), and the scattering approach to the Casimir effect is presented in \cite{scattering}.

The simplest theoretical way allowing to study the Casimir effect is the case of the interaction between two parallel plates placed in the vacuum. The quantum notion of vacuum corresponds to a collection of infinite set of waves which contemplate all wavelength possibilities; however, when two parallel plates are placed in this vacuum, only specific wavelengths are possible between them. 

In the usual Quantum Field Theory, the Lorentz symmetry is preserved. However,  the seminal paper \cite{Carrol_1990} called interest to the possibility of its breaking. The Lorentz-breaking theories have attracted a great deal of attention during the recent years,  it is worth to mention the Lorentz-breaking extension of the Standard Model \cite{Kostelecky}. One of the most important directions in their study is the investigation of possible impacts of  Lorentz-violating additive terms for field theory models both at classical and quantum levels.  We note that this approach is based on introducing the explicit Lorentz symmetry breaking in the action. This approach should be distinguished from introducing the Lorentz-breaking boundary conditions {\it ad hoc} for the Lorentz-invariant Lagrangian, since in the case of the Lorentz-breaking terms of the Lagrangian, the existence of privileged space-time directions is an intrinsic feature of the theory, while the Lorentz-breaking boundary conditions affect only the given physical situation. A large list of studies of Lorentz-breaking boundary conditions within the Casimir effect for the spinor field is presented in \cite{spinfield}, and for the scalar field in \cite{scalfield}.

The  study of the Casimir effect for scalar quantum fields confined between two large parallel plates in the presence of Lorentz symmetry breaking has been developed in \cite{Cruz_2017}. Moreover, the thermal corrections for the corresponding Casimir energies  have been investigated in \cite{Cruz_2018}. In both papers, it was supposed that the scalar field obeys different boundary conditions on the plates and that the Lorentz symmetry breaking is introduced through contracting of the derivative of this field with a constant vector, $u^\mu$. Also, the Horava-Lifshitz approach \cite{HL} breaks the Lorentz symmetry, but unlike the methodology proposed in \cite{Kostelecky}, in this case the Lorentz symmetry breaking is strong, being introduced through an explicit asymmetry between space and time coordinates (space-time anisotropy).  Within this approach,  the Casimir energy and pressure have been calculated in \cite{Petrov,Icaro}. 

In this paper we would like to  carry out the analysis of Casimir energy and pressure, considering at this time a massive quantum fermionic fields confined between two parallel plates in a Lorentz-violating model. In order to prevent crossing of plates by the fermionic current, we adopt the MIT bag model \cite{Ken}. The analysis of Casimir energy, in a Lorentz-symmetric theory, for the case of the massive fermionic quantum field  confined between two parallel plates in a high-dimensional spacetime have been developed in \cite{Elizalde_2003}, and in \cite{Aram_2009} such a study has been done for a non-trivial topology for the space section. 

This paper is organized as follows. In section \ref{Sec2} we present the model of QFT with Lorentz symmetry violation for fermionic fields, admitting a direct coupling between an arbitrary constant vector with the derivative of the field. In section \ref{Sec3}, we calculate the Casimir energy and pressure considering the field obeying the MIT bag model boundary condition on two parallel plates, considering different directions for the arbitrary constant vector. Our most relevant results are summarized in section \ref{Concl}. In appendix \ref{App} we present our proposal of the modified Lagrangian for the MIT bag model. Here, we assume $\hbar = c = 1$, and the metric signature will be taken as $(+,-,-,-)$.


\section{The Lorentz violating model}
\label{Sec2}

In this section, we introduce the theoretical model which we want to investigate. In fact, this model has been proposed in \cite{carroll2008aether}, and is described by  a massive fermionic quantum field characterized by the Lagrangian density below:
\begin{eqnarray}
 \label{lagrangian_density}
  \mathcal{L} = \bar{\Psi}(x) \big{(} i \gamma^{\mu} \partial_{\mu} - m + i \lambda u^{\mu} u^{\nu} \gamma_{\mu} \partial_{\nu} \big{)} \Psi(x) \ .
\end{eqnarray}
In the above equation, the dimensionless parameter $\lambda$ is supposed to be much smaller than unity. It  measures intensity of the Lorentz symmetry violation of the system caused by the presence of a coupling between the derivative of the field with an arbitrary constant vector $u^\mu$  defined so that $u^{\mu}u_{\mu}$ can be equal to $1,-1$ or $0$. This vector introduces the privileged space-time direction. In a certain sense, the theory is characterized by four constant parameters: $\lambda$ characterizing intensity of Lorentz symmetry breaking and three independent components of $u_{\mu}$ characterizing its direction.

Here, we will assume that the Dirac matrices $\gamma^{\mu}$ are given in the following representation:
\begin{eqnarray}
 \label{Dirac_matrix}
 \begin{aligned}
  \gamma^0 = \begin{pmatrix}
    I & 0 \\
    0 & -I
\end{pmatrix} , \ \ \ \ \ \ \ \ \ \ \ 
\gamma^j = \begin{pmatrix}
    0 & \sigma^j \\
    -\sigma^j & 0
\end{pmatrix} ,
 \end{aligned}
\end{eqnarray}
where $I$ represents the $2\times2$ identity matrix, and $\sigma^j$ the Pauli matrices.

The modified Dirac equation reads,
\begin{eqnarray}
 \label{Dirac_equation}
  \big{(} i \gamma^{\mu} \partial_{\mu} - m + i \lambda u^{\mu} u^{\nu} \gamma_{\mu} \partial_{\nu} \big{)} \Psi(x) = 0 \ .
\end{eqnarray}
The corresponding energy-momentum tensor is given by
\begin{eqnarray}
 \label{energy_tensor}
  T^{\mu \nu} = \bar{\Psi}(x) \big{(} i \gamma^{\mu} + i \lambda u^{\mu} u^{\alpha} \gamma_{\alpha} \big{)} \partial^{\nu} \Psi(x)  \ . 
\end{eqnarray}
We can verify that
\begin{eqnarray}
 \begin{aligned}
  T^{\mu \nu} - T^{\nu \mu} = i\lambda u^{\alpha} \bar{\Psi}(x) \gamma_{\alpha} \big{(}u^{\mu} \partial^{\nu}-u^{\nu} \partial^{\mu}\big{)} \Psi(x) ,
 \end{aligned}
\end{eqnarray}
where this asymmetry is a typical characteristic for theories with a Lorentz symmetry breaking. Moreover, we can verify that
\begin{eqnarray}
\partial_\mu T^{\mu\nu}=0  \  .
\end{eqnarray}


\section{The influence of Lorentz symmetry breaking in Casimir effect}
\label{Sec3}

\begin{figure}[!htb]
\centering
\includegraphics[scale=1.0]{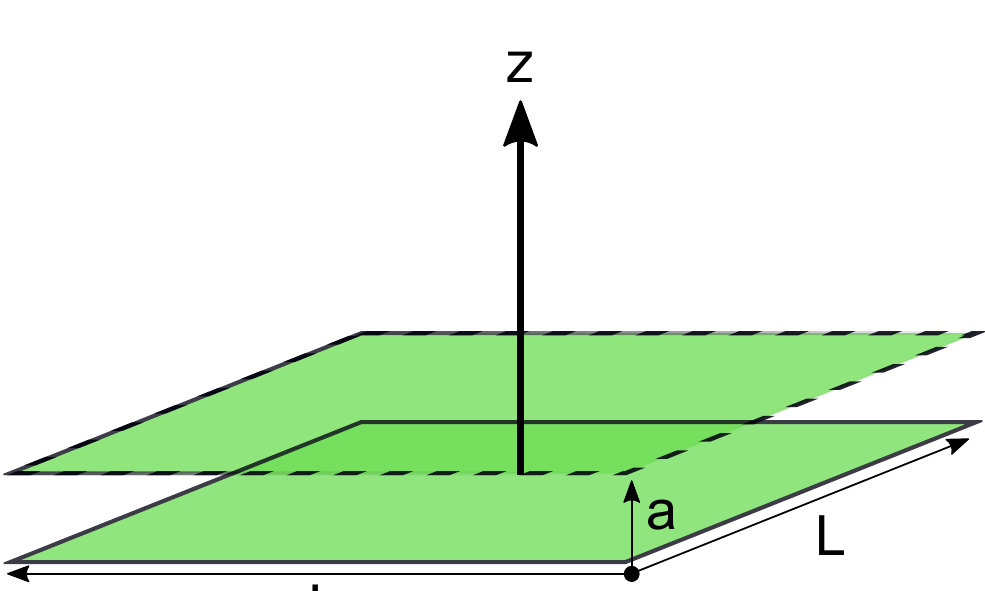}
\caption{Two parallel plates with area $L^{2}$ separated by a distance $a \ll L$.}
\label{fig_plates}
\end{figure}

In this section, we will analyze the influence of the space-time anisotropy, generated by the Lorentz-breaking term, on the Casimir energy associated with a massive fermionic quantum field confined between two large parallel plates. In this way, we assume two parallel plates placed at $z = 0$ and $z = a$, as is shown in Fig. \ref{fig_plates}.

Within this study, we will consider the cases where the constant $4$-vector, $u^{\mu}$, is either timelike or spacelike. For each case, we  will obtain the solution for the Dirac equation Eq. \eqref{Dirac_equation} imposing that the field obeys the MIT bag boundary condition below,
\begin{eqnarray}
 \label{MIT_bag_boundary}
 \big{(}I + i \gamma^{\mu} n_{\mu}\big{)} \Psi(x) \Big{|}_{z=0, a} = 0 \ ,
\end{eqnarray}
where $n_{\mu}$ is unit vector normal outgoing to the boundaries, i.e., in our case normal to the plates. For our configuration of plates we have consider $n_{\mu} = - \delta_{\mu}^z$ for the plate at $z = 0$ and $n_{\mu} = \delta_{\mu}^z$  at $z = a$.

\subsection{Timelike vector case}

In this subsection, we consider the case in that $4$-vector $u^{\mu}$ is timelike, i.e.
\begin{eqnarray}
 \label{timelike_4-vector}
 u^{(t)} = (1,0,0,0) .
\end{eqnarray}
Assuming the time dependence for the fermionic wave function in the form $e^{\mp i \omega t}$, for positive- and negative-energy solutions of equation \eqref{Dirac_equation} can be presented as
\begin{eqnarray}
 \label{solution_timelike}
 \begin{aligned}
  \Psi^{(+)}_{\beta}(x) = A_{\beta} e^{-i \omega t} \begin{pmatrix} \varphi(\vec{x}) \\ \frac{-i \sigma^j \partial_j \varphi(\vec{x})}{(1+\lambda)\omega + m}
\end{pmatrix} , \ \ \ \ \ 
\Psi^{(-)}_{\beta}(x) = A_{\beta} e^{i \omega t} \begin{pmatrix} \frac{i\sigma^j\partial_j \chi(\vec{x})}{(1+\lambda)\omega+m} \\ \chi(\vec{x})
\end{pmatrix} \  ,
 \end{aligned}
\end{eqnarray}
with the dispersion relation
\begin{eqnarray}
 \label{timelike_dispersion_relation}
 \omega =(1+\lambda)^{-1} \sqrt{k_x^2+k_y^2+k_z^2+m^2} \  ,
\end{eqnarray}
and the two-component spinors $\varphi$ and $\chi$ correspond to particles and antiparticles respectively and have the following explicit form:
\begin{eqnarray}
 \label{timelike_bispinors}
 \begin{aligned}
  \varphi(\vec{x}) = e^{i(k_x x + k_y y)} \Big{(}\varphi_+ e^{ik_z z} + \varphi_- e^{-ik_z z}\Big{)} ,  \ \
  \chi(\vec{x}) = e^{-i(k_x x + k_y y)} \Big{(}\chi_+ e^{ik_z z} + \chi_- e^{-ik_z z}\Big{)} \  .
 \end{aligned}
\end{eqnarray}
These fermionic modes are characterized by the set of quantum numbers $\{k_x,k_y,k_z\}$,  with $k_x,k_y$ being continuous quantum numbers, and $k_z$ must be discrete, to satisfy the boundary conditions. To be more precise, we will introduce a fourth quantum number.

From the boundary condition \eqref{MIT_bag_boundary} on the plate at $z = 0$ we find the following relations between the spinors in  \eqref{timelike_bispinors}:
\begin{eqnarray}
 \label{timelike_spinor_1}
 \begin{aligned}
  \varphi_+ = - \frac{m[(1+\lambda)\omega + m] + k_z^2 - k_z \sigma^3(\sigma^1 k_x+\sigma^2 k_y)}{(m-ik_z)[(1+\lambda)\omega + m]} \varphi_- , \\
  \chi_- = - \frac{m[(1+\lambda)\omega+m]+k_z^2-k_z\sigma^3(\sigma^1k_x+\sigma^2k_y)}{(m+ik_z)[(1+\lambda)\omega+m]} \chi_+ .
 \end{aligned}
\end{eqnarray}
Here it should be noted that, using (\ref{timelike_dispersion_relation}), one can immediately see that actually  these spinors do not depend on $\lambda$.

Substituting into \eqref{MIT_bag_boundary} at $z=a$, the fermionic functions, Eq.s \eqref{solution_timelike}, \eqref{timelike_bispinors}, followed by \eqref{timelike_spinor_1}, and after some intermediate steps, the transcendental equation below is obtained,
\begin{eqnarray}
\label{timelike_transcendental_equation}
\frac{a m}{k_z a} \sin(k_z a) + \cos(k_z a) = 0  \ .
\end{eqnarray}

The vacuum energy is obtained by taking the vacuum expectation value of $\hat{H}$:
\begin{eqnarray}
 \label{timelike_vacuum energy_1}
  E_0 &=& \langle 0|\hat{H}|0 \rangle = - \frac{(1+\lambda)L^2}{2\pi^2} \int d\vec{k} \sum_{n=1}^{\infty} \omega \nonumber\\
  &=& - \frac{L^2}{2\pi^2} \int_{-\infty}^{\infty} dk_x \int_{-\infty}^{\infty} dk_y \sum_{n=1}^{\infty} \sqrt{k_x^2+k_y^2+\Big{(}\frac{\alpha_n}{a}\Big{)}^2+m^2} ,
\end{eqnarray}
where we have defined $\alpha_n = k_n a$, being $k_n= k_z$, the order $n$ root of \eqref{timelike_transcendental_equation}. By the above expression, we clearly observe that the vacuum energy  does not depend on the parameter $\lambda$  parametrizing the intensity of the Lorentz symmetry violation. So, we stop this analysis at this point and go directly to the next cases.

\subsection{Spacelike vector case}

Here, we have three different directions possible for the 4-vector $u^{(\mu)}$, that is, $u^{(x)}=(0,1,0,0)$; $u^{(y)}=(0,0,1,0)$, and $u^{(z)}=(0,0,0,1)$. For the two first vectors, the Casimir energy and pressure are the same. So, let us consider only the case $u^{(x)}$:
\begin{eqnarray}
 \label{spacelike_vector}
 u^{(x)} = (0,1,0,0) .
\end{eqnarray}

By considering this constant vector, the solutions for \eqref{Dirac_equation} reads,
\begin{eqnarray}
 \label{solution_spacelike_x}
 \begin{aligned}
  \Psi^{(+)}_{\beta} = A_{\beta} e^{-i \omega t} \begin{pmatrix} \varphi(\vec{x}) \\ \frac{-i \sigma^j \partial_j \varphi(\vec{x})+i\lambda \sigma^1\partial_x \varphi(\vec{x})}{\omega + m}
\end{pmatrix} \ , 
\Psi^{(-)}_{\beta} = A_{\beta} e^{i \omega t} \begin{pmatrix} \frac{i\sigma^j\partial_j\chi(\vec{x})-i\lambda \sigma^1\partial_x \chi(\vec{x})}{\omega+m} \\ \chi(\vec{x})
\end{pmatrix}  \  , \\
 \end{aligned}
\end{eqnarray}
for positive- negative-energy wave-functions, where now the dispersion relation is
\begin{eqnarray}
 \label{spacelike_dispersion_relation_x}
 \omega = \sqrt{(1-\lambda)^2 k_x^2 + k_y^2 + k_z^2 + m^2} \  .
\end{eqnarray}
The two-component spinor reads,
\begin{eqnarray}
 \label{solution_timelike_functions}
 \begin{aligned}
  \varphi(x)= e^{i(k_x x + k_y y)} \Big{(}\varphi_+ e^{ik_z z} + \varphi_- e^{-ik_z z}\Big{)} , \  \newline
  \chi(x) = e^{-i(k_xx+k_yy)}\Big{(}\chi_+ e^{ik_zz}+\chi_- e^{-ik_zz}\Big{)}  .
 \end{aligned}
\end{eqnarray}
From the boundary condition \eqref{MIT_bag_boundary}  on the plate at $z = 0$, we get
\begin{eqnarray}
 \label{spacelike_spinor_x_1}
 \begin{aligned}
  \varphi_+ = - \frac{m(\omega+m)+k_z^2-k_z\sigma^3[(1-\lambda)\sigma^1 k_x+\sigma^2k_y]}{(m-ik_z)(\omega+m)} \varphi_- ,
 \end{aligned}
\end{eqnarray}
and
\begin{eqnarray}
 \label{spacelike_spinor_x_2}
 \begin{aligned}
  \chi_- = - \frac{m(\omega+m)+k_z^2-k_z\sigma^3[(1-\lambda)\sigma^1k_x+\sigma^2k_y]}{(m+ik_z)(\omega+m)} \chi_+.
 \end{aligned}
\end{eqnarray}
Moreover, from the boundary condition at $z = a$ follows the transcendental equation \eqref{timelike_transcendental_equation}.

For this case the vacuum energy is given by,
\begin{eqnarray}
 \label{spacelike_vacuum_energy_x}
 \begin{aligned}
  E_0 = \langle 0|\hat{H}|0\rangle = - \frac{L^2}{2\pi^2} \int d\vec{k} \sum_{n=1}^{\infty} \sqrt{(1-\lambda)^2 k_x^2+k_y^2+\Big{(}\frac{\alpha_n}{a}\Big{)}^2+m^2} .
 \end{aligned}
\end{eqnarray}
By using the coordinate transformation $\tilde{k}_x = (1-\lambda)k_x$, we get
\begin{eqnarray}
 \label{spacelike_vacuum_energy_x_2}
 \begin{aligned}
  E_0 = - \frac{L^2 (1-\lambda)^{-1}}{2\pi^2} \int_{-\infty}^{\infty}d\tilde{k}_x \int_{-\infty}^{\infty}d k_y \sum_{n=1}^{\infty} \sqrt{\tilde{k}_x^2+k_y^2+\Big{(}\frac{\alpha_n}{a}\Big{)}^2+m^2} .
 \end{aligned}
\end{eqnarray}
So the vacuum energy, in this case, is the same that of the timelike case \eqref{timelike_vacuum energy_1} multiplied by the factor $(1-\lambda)^{-1}$.
Since in this case the Lorentz-breaking modification is trivial,  we will not  proceed further because this situation is a particular case of the next calculation.

Finally, let us consider the case where the $4$-vector is orthogonal to plates:
\begin{eqnarray}
 \label{spacelike_vector_z}
 u^{(z)} = (0,0,0,1)  \ .
\end{eqnarray}
Again we solve the Dirac equation Eq. \eqref{Dirac_equation}, assuming the time dependence in the form $e^{\mp i \omega t}$. The positive- and negative-energy solutions to this case can read
\begin{eqnarray}
 \label{spacelike_solutions_z}
 \begin{aligned}
  \Psi_{\beta}^{(+)} = A_{\beta} e^{-i \omega t} \begin{pmatrix} \varphi(\vec{x}) \\ \frac{-i \sigma^j \partial_j \varphi(\vec{x})+i \lambda \sigma^3 \partial_z \varphi(\vec{x})}{\omega + m}
\end{pmatrix}
, \Psi_{\beta}^{(-)} = A_{\beta} e^{i \omega t} \begin{pmatrix} \frac{i\sigma^j\partial_j \chi(\vec{x})-i\lambda \sigma^3\partial_z \chi(\vec{x})}{\omega+m} \\ \chi(\vec{x}) \end{pmatrix},
 \end{aligned}
\end{eqnarray}
where the dispersion relation is now given by
\begin{eqnarray}
 \label{spacelike_dispersion_relation_z}
 \omega = \sqrt{k_x^2+k_y^2+(1-\lambda)^2k_z^2+m^2} ,
\end{eqnarray}
and
\begin{eqnarray}
 \label{spacelike_spinor_z}
 \begin{aligned}
  \varphi(x)= e^{i(k_xx+k_yy)} \Big{(} \varphi_+ e^{ik_zz}+\varphi_- e^{-ik_zz} \Big{)} , \ \newline  \chi(x)= e^{-i(k_xx+k_yy)}\Big{(}\chi_+ e^{ik_zz}+\chi_- e^{-ik_zz} \Big{)} .
 \end{aligned}
\end{eqnarray}
From the boundary condition \eqref{MIT_bag_boundary} applied on the plate at $z = 0$, we get
\begin{eqnarray}
 \label{spacelike_spinor_z_1}
 \begin{aligned}
  \varphi_+ = - \frac{m(\omega+m)+(1-\lambda)^2k_z^2-(1-\lambda)k_z\sigma^3(\sigma^1k_x+\sigma^2k_y)}{[m-i(1-\lambda)k_z](\omega+m)} \varphi_- , \\ \chi_- = - \frac{m(\omega+m)+(1-\lambda)^2k_z^2-(1-\lambda)k_z\sigma^3(\sigma^1k_x+\sigma^2k_y)}{[m+i(1-\lambda)k_z](\omega+m)} \chi_+ .
 \end{aligned}
\end{eqnarray}
As to  the plate on $z=a$, we find
\begin{eqnarray}
 \label{spacelike__z_transcendental_equation}
 \begin{aligned}
  \frac{a m}{(1-\lambda)k_z a} \sin(k_z a) + \cos(k_z a) = 0  \ .
 \end{aligned}
\end{eqnarray}

The vacuum energy for this case reads
\begin{eqnarray}
 \label{spacelike_z_vacuum_energy_1}
 \begin{aligned}
  E_0 &= - \frac{L^2}{2\pi^2} \int_{-\infty}^{\infty} dk_x \int_{-\infty}^{\infty} dk_y \sum_{n=1}^{\infty} \sqrt{k_x^2+k_y^2+(1-\lambda)^2k_z^2+m^2} \  . 
 \end{aligned}
\end{eqnarray}
Again using the definition $\alpha_n = k_n a$, being $k_n=k_z$ the n-root of \eqref{spacelike__z_transcendental_equation} we get
\begin{eqnarray}
 \label{spacelike_z_vacuum_energy_2}
 \begin{aligned}
  E_0 &= - \frac{L^2}{2\pi^2} \int_{-\infty}^{\infty} dk_x \int_{-\infty}^{\infty} dk_y \sum_{n=1}^{\infty} \sqrt{k_x^2+k_y^2+\bigg{(}\frac{(1-\lambda)\alpha_n}{a}\bigg{)}^2+m^2}  \ .
 \end{aligned}
\end{eqnarray}
By the above expression, we clearly observe that in this case, the vacuum energy depends on the parameter describing the intensity of  the Lorentz violation in a non-trivial way. So, we will develop below the detailed analysis for this case.

As we can see, the vacuum energy is divergent. In order to  obtain a finite and well defined result for the Casimir energy, we have to renormalize it. This can be done by using the Abel-Plana summation formula below \cite{saharian2006generalized}:
\begin{eqnarray}
 \label{abel_plana_sum}
  \sum_{n=1}^{\infty} \frac{\pi f(\alpha_n)}{1-\frac{\sin(2\alpha_n)}{2\alpha_n}} = - \frac{\pi am f(0)}{2(am+1)} + \int_0^{\infty} dz f(z) - i\int_0^{\infty} dt \frac{f(it)-f(-it)}{\frac{t+am}{t-am}e^{2t}+1} \ .
\end{eqnarray}
To develop the summation over $n$ in \eqref{spacelike_z_vacuum_energy_2} we  must rewrite the integrand in an appropriate manner. The steps adopted from  now use the approach developed in \cite{Aram_2009}. So let  us start with the denominator appearing in the left hand side of the summation \eqref{abel_plana_sum}. By using \eqref{spacelike__z_transcendental_equation}, it is possible to obtain the identity below:
\begin{eqnarray}
 \label{timelike_relation_1}
 1 - \frac{\sin(2\alpha_n)}{2\alpha_n} = 1 + \frac{bm}{(bm)^2 + \alpha_n^2} \ .
\end{eqnarray}
So, we arrive at the regularized vacuum energy given by
\begin{eqnarray}
 \label{timelike_vacuum_energy_2}
 \begin{aligned}
  E_0 &= - \frac{L^2}{2\pi^3b} \int_{-\infty}^{\infty} dk_x \int_{-\infty}^{\infty} dk_y \bigg{(} - \frac{\pi bmf(0)}{2(bm+1)} + \int_0^{\infty} dz f(z)  \\ &- i \int_0^{\infty} dt \frac{f(it)-f(-it)}{\frac{t+bm}{t-bm}e^{2t}+1} \bigg{)} ,
 \end{aligned}
\end{eqnarray}
where the function $f(z)$ has been defined as
\begin{eqnarray}
 \label{timelike_function_f}
 f(z) = \sqrt{z^2 + \big{(}k_x^2 + k_y^2\big{)}b^2 + m^2b^2} \bigg{(}1 + \frac{bm}{(bm)^2 + z^2}\bigg{)} \  , 
\end{eqnarray}
with the $b$ parameter given by
\begin{eqnarray}
\label{b_rel}
 b=\frac a{(1-\lambda)}  \  .
\end{eqnarray}
The first term on the right-hand side of the equation \eqref{timelike_vacuum_energy_2} refers to vacuum energy in the presence of only one plate, and the second one is related with vacuum energy without plates. So, after renormalization, the Casimir energy is given by
\begin{eqnarray}
 \label{timelike_vacuum_energy_3}
 \begin{aligned}
  E_C &= \frac{iL^2b}{2\pi^3} \int_{-\infty}^{\infty} dk_x \int_{-\infty}^{\infty} dk_y \int_{0}^{\infty} du \frac{u-m}{(u+m)e^{2bu}+u-m} \bigg{(} 1 + \frac{m}{bm^2-bu^2}\bigg{)} \\ &\times \bigg{(}\sqrt{(iu)^2+k_x^2+k_y^2+m^2}-\sqrt{(-iu)^2+k_x^2+k_y^2+m^2}\bigg{)} ,
 \end{aligned}
\end{eqnarray}
where we have made the change of  variable $t=bu$. At this point we can separate the integral over the $u$ variable in two intervals: the first one is $\big{[}0, \ \sqrt{k_x^2+k_y^2+m^2}\big{]}$ and the second is $\big{[}\sqrt{k_x^2+k_y^2+m^2}, \ \infty \big{)}$. One finds that the integral in the first interval vanishes. So, we get the following Casimir energy
\begin{eqnarray}
 \label{timelike_vacuum_energy_4}
 \begin{aligned}
  E_C &= - \frac{L^2}{\pi^3} \int_{-\infty}^{\infty} dk_x \int_{-\infty}^{\infty} dk_y \int_{\sqrt{k_x^2+k_y^2+m^2}}^{\infty} du \
    \sqrt{u^2-k_x^2-k_y^2-m^2} \\ & \times \bigg{(}\frac{b(u-m)-m/(u+m)}{(u+m)e^{2bu}+u-m}\bigg{)} .
 \end{aligned}
\end{eqnarray}
Using the integral relation below \cite{Aram_2009}
\begin{eqnarray}
 \label{integral_relation_1}
 \begin{aligned}
  \int d\vec{k}_p \int_{\sqrt{\vec{k}_p^2+c^2}}^{\infty} dz (z^2 - \vec{k}_p^2 -c^2)^{\frac{s+1}{2}} f(z) = \frac{\pi^{\frac{p}{2}}\Gamma[\frac{s+3}{2}]}{\Gamma[\frac{p+s+3}{2}]} \int_c^{\infty} dx (x^2-c^2)^{\frac{p+s+1}{2}} f(x) ,
 \end{aligned}
\end{eqnarray}
together with the identification
\begin{eqnarray}
 \label{relation_identification_1}
 \bigg{(}\frac{b(x-m)-m/(x+m)}{(x+m)e^{2bx}+x-m}\bigg{)} = -\frac{1}{2} \frac{d}{dx} \ln \Big{(}1+\frac{x-m}{x+m}e^{-2bx}\Big{)} ,
\end{eqnarray}
we find the following expression to Casimir energy
\begin{eqnarray}
 \label{orthogonal_casimir_energy}
 E_C = - \frac{L^2}{\pi^2 b^3} \int_{0}^{\infty} dy\big{(}y+bm\big{)} \sqrt{y\big{(}y+2bm\big{)}} \ln \bigg{(}1+\frac{y}{y+2bm}e^{-2(y+bm)} \bigg{)} ,
\end{eqnarray}
where we have made the integration  by parts and  the changed the variable: $bx = y+bm$. Another procedure adopted to develop the summation over the $n$ in \eqref{spacelike_z_vacuum_energy_2}, is based on using Riemann zeta function regularization \cite{SH}.  Within this approach we  replace $\sqrt{k_x^2+k_y^2+(1-\lambda)^2k_n^2}$, in the integrand, by $[k_x^2+k_y^2+(1-\lambda)^2k_n^2]^{-s}$. Afterwards,  we  develop the integration over $k_x$ and $k_y$, and the remaining sum over $n$ converges for all complex value of $s$ for  sufficiently large value of $Re(s)$. Finally the Casimir energy can be obtained by evaluating the corresponding zeta function at the pole $s=-1/2$.

Unfortunately for massive fields there is no closed expression for the integral in \eqref{orthogonal_casimir_energy}. However we can obtain approximate results for two limiting cases: $am \ll 1$ and $am \gg 1$, corresponding to case of small distance between plates and/or small mass, and of large distance between plates and/or large mass, respectively:\\
$(i)$ For the case $am \ll 1$, we can expand the integrand in powers of $bm$. Up to the first order we get:
\begin{eqnarray}
{x}^{2}\ln  \left( 1+{{\rm e}^{-2\,x}} \right) - \left( 2\,{\frac {x{
{\rm e}^{-2\,x}} \left( x+1 \right) }{1+{{\rm e}^{-2\,x}}}}-2\,x\ln 
\left( 1+{{\rm e}^{-2\,x}} \right)  \right) bm \ .
\end{eqnarray}
Substituting the above expression in \eqref{orthogonal_casimir_energy}, we obtain
\begin{eqnarray}
\label{spacelike_z_casimir_energy_little_ma}
\begin{aligned}
\frac{E_C}{L^2} = - \frac{7\pi^2(1-\lambda)^3}{2880 a^3} \bigg{(}1-\frac{120 am}{7\pi^2 (1-\lambda)} \bigg{)} 
\end{aligned}
\end{eqnarray}
and consequently, the Casimir pressure  is given by
\begin{eqnarray} 
\label{spacelike_z_casimir_pressure}
P_C = -\frac{(1-\lambda)^2 \big{(}7\pi^2(1-\lambda)-80 am \big{)}}{960 a^4} \ .
\end{eqnarray}
$(ii)$ For the case $am \gg 1$, the most relevant contribution in the expansion of the integrand of \eqref{orthogonal_casimir_energy}, is:
\begin{eqnarray}
e^{-2(mb+x)}x^{3/2}\sqrt{\frac{b m}{2}} \ .
\end{eqnarray}
In this approximation the Casimir energy reads,
\begin{eqnarray}
\label{spacelike_z_casimir_energy_big_ma}
\begin{aligned}
\frac{E_C}{L^2} = - \frac{3 m^{1/2}(1-\lambda)^{5/2}}{32\pi^{3/2} a^{5/2}} e^{-\frac{2am}{1-\lambda}} \ .
\end{aligned}
\end{eqnarray}
We can see, that in this case, the Casimir energy per area unit decays exponentially with $a m$.

\begin{figure}[!htb]
\centering
\includegraphics[scale=1]{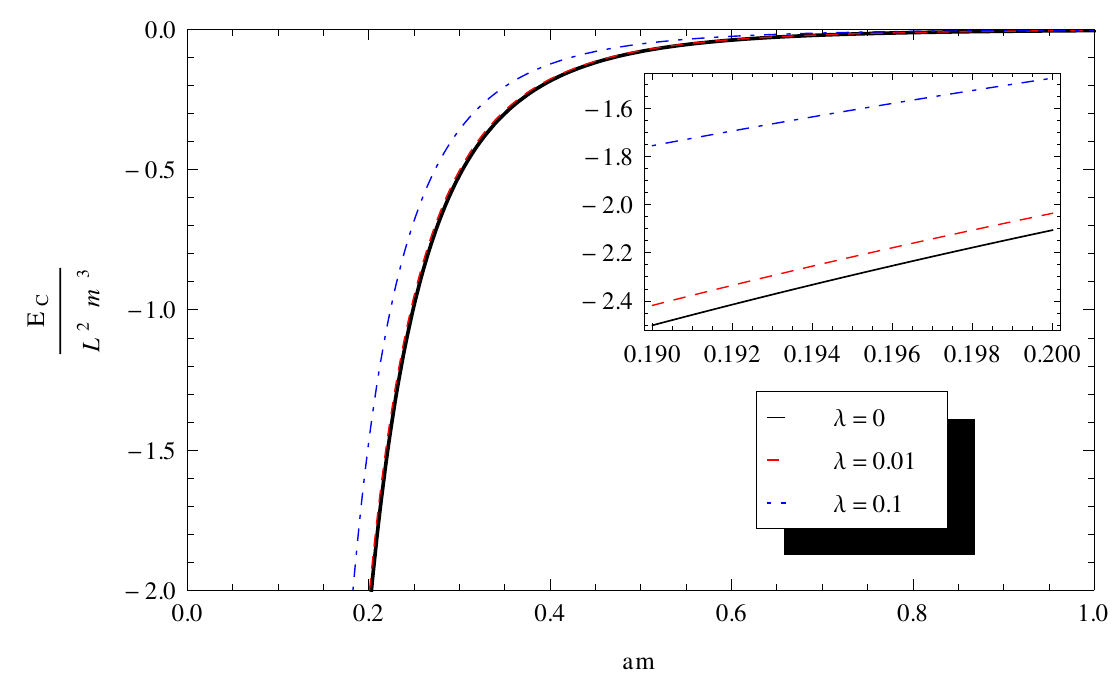}
\caption{Casimir energy density, for case $u^{(z)}=(0,0,0,1)$, in units of  $L^2m^3$, as a function of $a m$ for the values of parameter $\lambda = 0.0, 0.01, 0.1$ .}
\label{energy}
\end{figure}

\begin{figure}[!htb]
\centering
\includegraphics[scale=1]{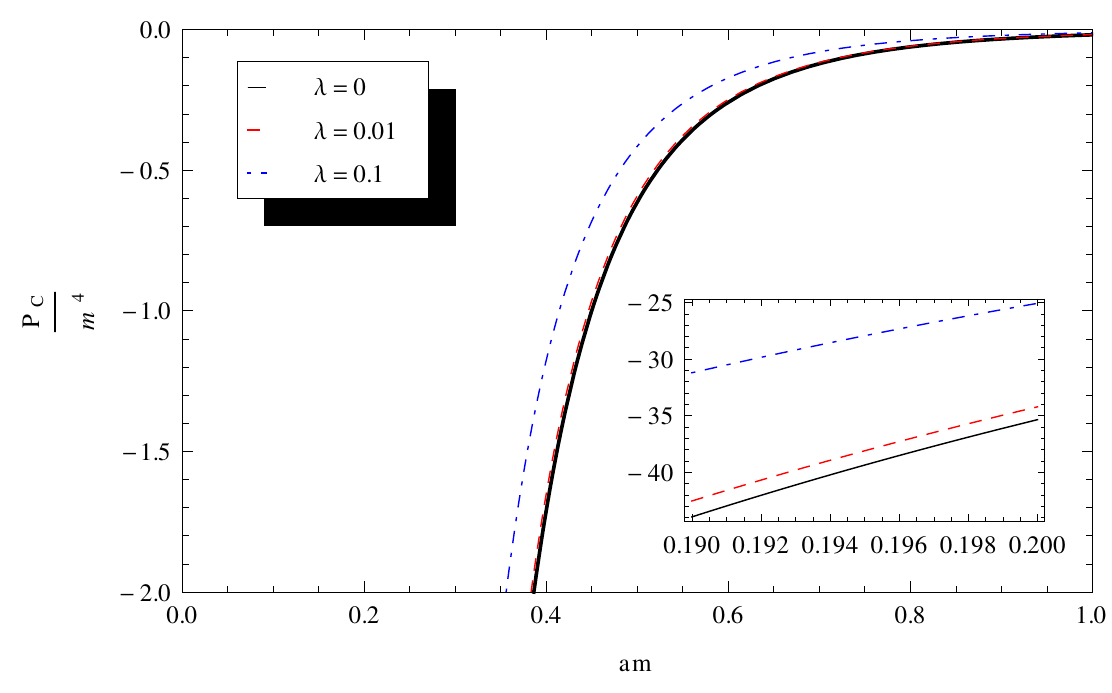}
\caption{Casimir pressure, for the case $u^{(z)}=(0,0,0,1)$, in units of $m^{4}$, as a function of $am$ for the values of parameter $\lambda = 0.0, 0.01, 0.1$ .}
\label{pressure}
\end{figure}

In figures \ref{energy} and \ref{pressure}, we exhibit the behavior of the Casimir energy, Eq. \eqref{orthogonal_casimir_energy}, and pressure, respectively, as a function of $a m$ considering different values of the parameter characterizing the intensity of the Lorentz symmetry violation, $\lambda$. By these plots we can see that both quantities depend on this parameter, in the sense that they increase when we increase $\lambda$.

In the publication \cite{Elizalde_2003}, the authors have obtained, in terms of a series expansion, the Casimir energy associated with fermionic fields obeying MIT boundary condition on two parallel plates of area $L^2$ and separated by a distance $a$. Their result is reproduced below,
\begin{eqnarray}
 \label{timelike_casimir_energy_2}
 E_C = - \frac{L^2 m^{1/2}}{4 \pi^2 a^{5/2}} \sum_{k=1}^{\infty} \frac{(-1)^{k+2}}{k^{7/2}} \Gamma \Big{(}{3}/{2}+k \Big{)}
 \frac{d}{d\beta} \Big{[} \beta^{-3/2} W_{-k,1}(4amk\beta) \Big{]} \Big{|}_{\beta=1} ,
\end{eqnarray}
where $W_{\mu,\nu}(z)$ represents the Whittaker function \cite{Abramo}.
By using the above expression the authors of \cite{Elizalde_2003} found approximate results for the Casimir energy for the limits $am \ll 1$ and $am \gg 1$. Their result for the limit $ma \ll 1$ disagrees with ours.

On  the base of the result \eqref{timelike_casimir_energy_2}, we can adapt this formula for our case. The corresponding expression is below:
\begin{eqnarray}
\begin{aligned}
E_C = - \frac{L^2 m^{1/2}}{4 \pi^2 b^{5/2}} \sum_{k=1}^{\infty} \frac{(-1)^{k+2}}{k^{7/2}}\Gamma \Big{(}{3}/{2}+k \Big{)} \frac{d}{d\beta}\Bigg{[}\beta^{-3/2} W_{-k,1}\Big{(}4b m k \beta \Big{)} \Bigg{]} \Bigg{|}_{\beta=1} \ .
\end{aligned}
\end{eqnarray}
By using \cite{Abramo} we write below the asymptotic expression for the  Whittaker function for small argument:
\begin{eqnarray}
\begin{aligned}
W_{\mu,\nu}(z) \approx \frac{\Gamma(2\nu)}{\Gamma(1/2-\mu+\nu)} \Bigg{(}z^{1/2-\nu}+\frac{\mu z^{3/2-\nu}}{2\nu-1}\Bigg{)} \ .
\end{aligned}
\end{eqnarray}
By using this expression we obtain \eqref{spacelike_z_casimir_energy_little_ma}.

\section{Concluding Remarks}
\label{Concl}
In this paper we have investigated the fermionic Casimir energy  for the aether-like CPT-even Lorentz-violating (LV) model, by admitting a direct coupling between the derivative of the  field and an arbitrary constant vector. We assume that the field is confined between two parallel plates by imposing the MIT bag model boundary condition on each plate. 

For each arbitrary direction of the constant vector, $u^{\mu}$, the Casimir energy has been analyzed. It was observed that the dispersion relations are modified by the presence of the parameter, $\lambda$, associated with the LV parameter, $\lambda$, as exhibited in \eqref{timelike_dispersion_relation}, \eqref{spacelike_dispersion_relation_x} and \eqref{spacelike_dispersion_relation_z}. However, due to the structure of the energy-momentum tensor, Eq. \eqref{energy_tensor}, there is no modification on the Casimir energy  in the case of the time-like vector; as to the constant vector parallel to the plates, the correction on the Casimir energy is given by a multiplicative factor, as can be seen in \eqref{spacelike_vacuum_energy_x_2}.  
The most relevant correction takes place for the case when the constant vector is perpendicular to the plates. In this case the correction associated with the Lorentz-violating parameter appears not only  as a multiplicative factor, but also in the integrand of the integral representation \eqref{orthogonal_casimir_energy}, through  $b=a/(1-\lambda)$. For this case we have developed in detail all the steps of calculation of the Casimir energy by using the Abel-Plana summation formula \eqref{abel_plana_sum}. Because the integral in \eqref{orthogonal_casimir_energy} cannot be evaluated in a closed form, we present the Casimir energy for the cases where $am \ll 1$, Eq. \eqref{spacelike_z_casimir_energy_little_ma},  and $am \gg 1$, Eq. \eqref{spacelike_z_casimir_energy_big_ma}. For the first case we also obtained the Casimir pressure, Eq. \eqref{spacelike_z_casimir_pressure}. Moreover, we presented in figures \ref{energy} and \ref{pressure} the behavior of Casimir energy in units of $L^2m$ and pressure in units of $m^4$, for different values of $\lambda$. We could observe that both quantities depend on this parameter, in the sense that they increase when we increase $\lambda$.

Our final remarks about this system concern the Hermiticity properties of the Hamiltonian associated with the Lagrangian density \eqref{lagrangian_density}. At the end of the Appendix we will argue that in fact the Hamiltonian modified due to the additive Lorentz-violating term is a self-adjoint operator.

{\bf Acknowledgements.} This work was partially supported by Conselho
Nacional de Desenvolvimento Cient\'{\i}fico e Tecnol\'{o}gico (CNPq). A. Yu. P. has been partially supported by the CNPq through the project No. 303783/2015-0, E. R. Bezerra de Mello through the project No. 313137/2014-5. M. B. Cruz has been supported by  Coordena\c{c}\~ao de Aperfei\c{c}oamento de Pessoal de N\'{i}vel Superior (CAPES).
\appendix
\section{Modified Lagrangian for the MIT bag model}
\label{App}

In this appendix  we propose the modified Lagrangian, i.e., in the presence of Lorentz violating term, for the MIT boundary condition. So, firstly we will express the density Lagrangian as being:
\begin{eqnarray}
\label{Lagrang}
\begin{aligned}
\mathcal{L} &= \Big{\{} \frac{i}{2}
\Big{[}\bar{\Psi} \gamma^{\mu} (\partial_{\mu}\Psi) - (\partial_{\mu}\bar{\Psi})\gamma^{\mu}\Psi\Big{]} - m\bar{\Psi}\Psi + \frac{i\lambda u^{\mu} u^{\nu}}{2}\Big{[}\bar{\Psi}\gamma_{\mu}(\partial_{\nu}\Psi) - (\partial_{\nu}\bar{\Psi})\gamma_{\mu}\Psi\Big{]} - B \Big{\}}\theta_v \\ &- \frac{1}{2} \Delta_s\bar{\Psi}\Psi [1+\lambda (u\cdot n)] \ ,
\end{aligned}
\end{eqnarray}
where there are explicit interactions between the constant vector $u^\mu$ with spinor field in the region inside the bag and on the bag itself, as explained in the following:\footnote{The factor $B$ in \eqref{Lagrang} is related with the energy-momentum tensor on the bag.} In \eqref{Lagrang} $\theta_v$ is the  Heaviside function that is equal to unity inside the bag and zero outside it, and $\Delta_s$ is the Dirac- delta function given by the spatial derivative of  $\theta_v$ as shown below,
\begin{eqnarray}
\frac{\partial\theta_v}{\partial x^\mu}=n_\mu\Delta_s \ , 
\end{eqnarray}
with $n_\mu$ is the unit vector normal to the surface. 

The equation of motion for this system is provided by the usual Euler-Lagrange formalism:
\begin{eqnarray}
\frac{\partial \mathcal{L}}{\partial \bar{\Psi}} - \partial_{\mu} \Big{[} \frac{\partial \mathcal{L}}{\partial (\partial_{\mu} \bar{\Psi})} \Big{]} = 0 \ .
\end{eqnarray}
Substituting \eqref{Lagrang} into the above equation, we obtain
\begin{eqnarray}
\label{EM1}
&&\Big{[}i\gamma^{\mu}(\partial_{\mu}\Psi) - m\Psi + i\lambda u^{\mu} u^{\nu} \gamma_{\mu} (\partial_{\nu}\Psi)\Big{]} \theta_v - \frac{i}{2} \Big{[} \Big{(}\gamma^{\nu} \Psi + \lambda u^{\mu} u^{\nu} \gamma_{\mu} \Psi \Big{)}n_{\nu}\nonumber\\
&&- i \Big{(}1+\lambda (u\cdot n)\Big{)} \Psi \Big{]} \Delta_s = 0 \ .
\end{eqnarray}

From \eqref{EM1}, two different equations are obtained:\\
$(i)$ In the region inside the bag, $\Delta_s=0$ and $\theta_v=1$, resulting in
\begin{eqnarray}
\left(i\gamma^{\mu}\partial_{\mu} - m + i\lambda u^{\mu} u^{\nu} \gamma_{\mu} \partial_{\nu} \right)\Psi = 0 \ .
\end{eqnarray}
$(ii)$ On the surface the bag, $\Delta_s=\infty$ and $\theta_v=0$, giving
\begin{eqnarray}
\left[(1+\lambda(u\cdot n))+i(1+\lambda(u\cdot n))(\gamma\cdot n)\right]\Psi = 0 \ ,
\end{eqnarray}
that coincides with the standard expression for the MIT bag model.

So we conclude that, in the above modified Lagrangian for the MIT bag model,  there is no influence of the LV parameter, $\lambda$, in the boundary condition imposed to the fermionic fields on the boundary.

An important point that deserves to be analyzed in this paper, and in some sense is connected with the subject treated in this Appendix, is related with the self-adjointness property of the Dirac Hamiltonian associated with \eqref{lagrangian_density}, when acts in the function that obey the MIT bag boundary condition \eqref{MIT_bag_boundary}. 
Modification of the Hamiltonian with the Lorentz-violating additive term, $i{\gamma}_{\mu}u^{\mu}u^{\nu}\partial_{\nu}$, in the case when the $u^{\mu}$ is parallel to one of the coordinate axes (just this case we considered throughout the paper),  rescales the derivative along this axis with a real factor. So, it does not affect the Hermiticity of the operator.


\end{document}